\input harvmac.tex
\input tables.tex
\def\KB{\overline{K}^0}
\def\PT{\widetilde P}
\def\singlespace{\baselineskip 12 pt}

\hyphenation{re-so-nan-ces}
\hbox to 16.5truecm{\hfil ROMA1 -- 1130/96}
\hbox to 16.5truecm{\hfil NAPOLI -- DSF-T-2/96}
\hbox to 16.5truecm{\hfil hep-ph/9601343}
\vskip 1.5 cm
\centerline{\titlerm {\bf Charm nonleptonic decays and final state interactions}}
\vskip 1.5 cm
\centerline{F. Buccella}
\vskip 10pt
\centerline{
{\it Dipartimento di Scienze Fisiche, Universit\`a di Napoli,
Napoli, Italy}}
\vskip 15pt
\centerline{M. Lusignoli$^{(*)}$ and
 A. Pugliese$^{(*)}$}
\vskip 10pt
\centerline{
{\it Dipartimento di Fisica, Universit\`a ``La Sapienza'', Roma,}}
\centerline{{\it and INFN, Sezione di Roma I, Roma, Italy}}
\vskip 2.5 truecm
\centerline {ABSTRACT}
\vskip 10truept
\singlespace\noindent
A global previous analysis of two-body nonleptonic decays of
$D$ mesons has been extended to the decays involving light scalar mesons.
The allowance for 
final state interaction also in nonresonant channels provides a fit of  
much improved quality and with less symmetry breaking in the axial charges.
We give predictions for about 50 decay branching ratios yet to be measured.
We also discuss long distance contributions to the difference 
$\Delta \Gamma$ between the $D_S$ and $D_L$ widths.
\vskip 2 cm
\noindent
$^{(*)}$ partially supported by the European Community under the Human
Capital and Mobility Programme, contract CHRX-CT93-0132.
\vfill\eject
\baselineskip 18 truept
\parindent=1cm
A theoretical description of exclusive nonleptonic decays of charmed hadrons
based on general principles is not yet possible. Even if the short distance
effects due to hard gluon exchange can be resummed and the effective
hamiltonian has been constructed at next-to-leading order
\ref\rHEFNL{G. Altarelli, G. Curci, G. Martinelli and S. Petrarca,
Nucl. Phys. B 187 (1981) 461 \semi A.J. Buras, M. Jamin, M.E. Lautenbacher
and P.E. Weisz, Nucl.Phys. B 370 (1992) 69 and Nucl.Phys. B 375 (1992) 501 
(addendum) \semi 
M. Ciuchini, E. Franco, G. Martinelli and L. Reina, 
Nucl. Phys. B 415 (1994) 403.}
, the evaluation of its matrix elements requires
nonperturbative techniques. A classic analysis based on QCD sum rules has been
presented in three papers by Blok and Shifman 
\ref\rBLOK{B. Blok and M. Shifman, Sov.Jour.Nucl.Phys. 45 (1987)
pp. 135, 301, 522.},
but only the general trends were reproduced: the agreement with
present data is poor at a quantitative level. 
Waiting for future progress in lattice QCD
calculations one has to rely on approximate methods and on models.

We recently presented 
\ref\rNOI{F. Buccella, M. Lusignoli, G. Miele, A. Pugliese and P. Santorelli,
Phys. Rev. D 51 (1995) 3478.} 
one such model,  
based on the factorized approximation, with annihilation terms and rescattering
effects due to resonances coupled to the final states, 
that has been rather 
successful in describing the bulk of the experimental 
data. Its main shortcoming was, in our opinion, the large flavour SU(3)
breaking in the axial charges forced by the fitting of the data on 
decay rates to final states with one pseudoscalar and one 
vector meson ($PV$).
 
In this letter we modify the previous approach, inserting rescattering
corrections also in nonresonant channels. Moreover, we include the
decays to final states containing one of the lowest mass scalar mesons ($S$), 
$f_0(980)$ and $a_0(980)$, that are connected through rescattering effects
to the previosly considered $PV$ final states. In this way, we are able
to obtain a much better fit of the experimental data, while keeping the
SU(3) breaking in the axial charges at a smaller and more acceptable
level. 

The scattering phase shifts in nonresonant channels were neglected
in \rNOI . For decays to $PP$ final
states this is essentially correct, given that only one nonresonating 
phase, corresponding to the  
$\underline{27}$ representation, is involved and of course the final rates
only depend on the phase shift differences. In the case of $PV$ final
states, on the other hand, many different SU(3) representations are present:
to minimize the number of parameters, we only include a nonzero phase shift 
for the 
$\underline{27}$ (besides the resonant $\underline{8}_F$) and keep the others
to zero. The $\underline{27}$ phase shift is most welcome, especially to obtain
a better fit for $D^+\to PV$ Cabibbo-allowed decays. We admit two different
values for the phase shift at the different energies, corresponding to
the masses of $D$ and of $D_s$: the fitted values 
are $\delta_{27}({\rm m}_D) = 47.4^\circ$ and 
 $\delta_{27}({\rm m}_{D_s}) = 59^\circ$, reasonably similar to each other, 
although maybe larger than expected.

The nature of the scalar resonances, $f_0(980)$ and $a_0(980)$, has been
discussed for quite a long time. They do not look like the members of a 
normal nonet, in that the $f_0$ is strongly coupled to $K\bar{K}$, and could be 
for this reason identified with an $s\bar{s}$ state, but it is degenerate
in mass with the isovector $a_0$. Moreover, the strange scalar states lie
quite a bit higher. For these reasons, it has been suggested that the
$f_0$ and $a_0$ are essentially 
$K\bar{K}$ molecules and are made therefore of two quarks and two antiquarks,
{\it i.e.} an $s\bar{s}$ pair plus a light $q\bar{q}$ pair 
\ref\rSCALAR{J. Weinstein and N. Isgur, Phys. Rev. Lett. 48 (1982) 659 \semi
J. Weinstein and N. Isgur, Phys. Rev. D 27 (1983) 588 \semi
N.N. Achasov and G.N. Shestakov, Zeits. f. Phys. C 41 (1988) 309.}, 
\ref\rISGUR{J. Weinstein and N. Isgur, Phys.Rev. D 41 (1990) 2236.}.
In charmed meson decays, both $D_s^+ \to f_0 \pi^+$ and $D^0 \to f_0 K_S$ 
have been observed experimentally 
\ref\rPDG{Review of Particle Properties, Particle Data Group,
Phys. Rev. D 50 (1994) part I.},
\ref\rBELLINI{E691 Collaboration, J.C. Anjos {\it et al.}, 
Phys.Rev.Lett. 62 (1989) 125 \semi
E687 Collaboration, P.L. Frabetti {\it et al.},  
Phys.Lett. B 351 (1995) 591 \semi
L. Moroni (E687), contribution to LISHEP-95 International School.},
\ref\rARGUS{ARGUS Collaboration, H. Albrecht {\it et al.},
Phys.Lett. B 308 (1993) 435 \semi
E687 Collaboration, P.L. Frabetti {\it et al.}, Phys.Lett. B 331 (1994)
217.}.
In the factorized approximation, one would have for the decay amplitudes
prior to rescattering corrections:

\eqn\eFACTOR{\eqalign{
{\cal A}_{\rm w}(D_s^+ \to f_0 \pi^+) &= \cr
     = - {G_F \over \sqrt 2}\,U_{ud}\,U_{cs}^*
  &\phantom{=\,}
  (C_2+\xi\,C_1)\;<f_0 |(A_s^c)_\mu|D_s^+>\;<\pi^+|(A_u^d)^\mu|0> \cr
  =  - {G_F \over \sqrt 2}\,U_{ud}\,U_{cs}^*
 &\phantom{=\,}
   (C_2+\xi\,C_1)\;f_\pi\;<f_0|\partial^\mu(A_s^c)_\mu|D_s^+>, \cr
{\cal A}_{\rm w}(D^0 \to f_0 \KB) &= \cr
     = - {G_F \over \sqrt 2}\,U_{ud}\,U_{cs}^*
  &\phantom{=\,}
   (C_1+\xi\,C_2)\;<f_0 |(A_u^c)_\mu|D^0>\;<\KB|(A_s^d)^\mu|0> \cr
  =  - {G_F \over \sqrt 2}\,U_{ud}\,U_{cs}^*
 &\phantom{=\,}
   (C_1+\xi\,C_2)\;f_K\;<f_0|\partial^\mu(A_u^c)_\mu|D^0>.\cr}}
In \eFACTOR ~$C_i$ are the Wilson coefficients in the effective 
hamiltonian, 
$\xi$ is the color screening parameter (that should be equal to $1/N_c$ if 
the factorization approach were exact), the axial currents are denoted by  
$(A_{q^{'}}^q)^{\mu} \equiv \bar{q}'\,\gamma^{\mu}\gamma_5\,q$ 
and we neglected possible annihilation contributions.
The observation of the decay $D_s^+\to f_0 \pi^+$ would imply the $s\bar{s}$
nature for $f_0$, while $D^0\to f_0 \KB$ points to a nonstrange 
composition.

Following the suggestion of \rISGUR ~we consider $f_0$ and $a_0$ as
cryptoexotic two--quark plus two-antiquark states and 
attribute them
to (incomplete) $\underline{8}$ and $\underline{1}$ SU(3) representations 
$|a_0\rangle \in |\underline{8}\rangle $, $|f_0\rangle \in \sqrt{1 \over 3}\,
|\underline{8}\rangle +\sqrt{2 \over 3}\, |\underline{1}\rangle $. 
We then define
\eqn\eAXIAL{\eqalign{
\langle f_0 | \partial^\mu (A_s^c)_\mu |D_s^+ \rangle =&
  (M^2_{D_s} - M^2_{f_0}) {a_{S} \over (1 - q^2/ M^2_{D_s})} \;, \cr
\langle f_0 | \partial^\mu (A_u^c)_\mu |D^0 \rangle =&
  (M^2_D - M^2_{f_0}) {a_{S} \over \sqrt{2} \,(1 - q^2/ M^2_D)} \;.}}
The axial charge $a_S$ is a parameter to be fitted. The result is 
$a_S\simeq0.39$, smaller -- as expected --
than the corresponding axial charges for $D$ transitions to vector mesons. 

 Our model also predicts charmed meson decays
to states including the $a_0(980)$ meson and Cabibbo suppressed decays to $PS$,
 not yet observed. The amplitudes in factorized
approximation are easily obtained, and the relevant form factors are all 
expressed in terms of the parameter $a_S$: as an example,
$$ \langle a_0^0 |\partial^\mu (A_d^c)_\mu |D^+ \rangle =
 -\;(M^2_D - M^2_{a_0}) {a_{S} \over \sqrt{2} (1 - q^2/ M^2_D)}\;.\eqno(2')$$

 We describe now in more detail the procedure followed to include
final state interactions. Defining as ${\cal B}$ the decay amplitude 
including the phase space factor, {\it i.e.} 
${\cal B}_{\rm w}={\cal A}_{\rm w}\;\sqrt{p\;/\;(8\,\pi\,m^2_D)}$
where $p$ is the momentum of the final particles in the $D$ rest frame,
we have for $D \to PV\;(PS)$ decays 
\eqn\eRESCA{\eqalign{
{\cal B} (D \to V_h\,P_k) = {\cal B}_{\rm w}
  (D \to V_h\,P_k) +& c_{hk}[\exp(i\delta_8)-1]
  A^8_T\;
  + d_{hk}[\exp(i\delta_{27})-1]
  A^{27}_T\;,\cr
{\cal B} (D \to S_h\,P_k) = {\cal B}_{\rm w}
  (D \to S_h\,P_k) +&x_{PS}\, \tilde
  {c}_{hk}[\exp(i\delta_8)-1]\;
  A^8_T\;+\cr
  +&y_{PS}\, \tilde{d}_{hk}[\exp(i\delta_{27})-1]\;
  A^{27}_T\;\ \ ,}}
where
\eqn\eTOTAL{\eqalign{
A^8_T =& {\sum_{h' k'} c_{h' k'}\,{\cal B}_{\rm w}
  (D \to V_{h'}\,P_{k'}) + x_{PS}
\sum_{h'' k''} \tilde{c}_{h'' k''}\,{\cal B}_{\rm w}
  (D \to S_{h''}\,P_{k''})  \over
\sum_{h' k'} |c_{h' k'}|^2  +
x_{PS}^2 \sum_{h'' k''} |\tilde{c}_{h'' k''}|^2 }\ ,\cr
A^{27}_T =& {\sum_{h' k'} d_{h' k'}\,{\cal B}_{\rm w}
  (D \to V_{h'}\,P_{k'}) + y_{PS}
\sum_{h'' k''} \tilde{d}_{h'' k''}\,{\cal B}_{\rm w}
  (D \to S_{h''}\,P_{k''})  \over
\sum_{h' k'} |d_{h' k'}|^2  +
y_{PS}^2 \sum_{h'' k''} |\tilde{d}_{h'' k''}|^2 }\ .}}

In \eRESCA ~and \eTOTAL ~$c_{hk}$ ($d_{hk}$) are the $P\,V$ 
couplings to $\underline{8}_F\;(\underline{27})$, multiplied by a
$(p / M_\rho)^{3\over2}$ kinematical
factor. This $p$ dependence must be present in the ${\cal B}$ amplitudes and, 
as in \rNOI , we include 
it in the coefficients in order to 
automatically decouple the channels below threshold. The $P\,S$
couplings
\foot{For the couplings of the singlet parts of $f_0$, $\eta$ and 
$\eta^{'}$ we adopt nonet symmetry.}
to $\underline{8}_D\;(\underline{27})$, multiplied by their 
kinematical factor (in this case $(p / M_\rho)^{1\over2}$), 
are denoted $\tilde{c}_{hk}$ ($\tilde{d}_{hk}$).  

We note that the phase shift $\delta_8$ is determined by the parameters of
the resonance $\PT$ appropriate to the decay channel considered
($\PT$ = $K(1830)$ or $\pi(1770)$), as follows
\eqn\ePHSH{\sin \delta_8 \, \exp (i \delta_8) =
   {\Gamma ({\PT} ) \over 2\,(M_{\PT} - M_D) - i\,\Gamma ({\PT} )}\;.}
In the isoscalar case, $\delta_8^{I=0}$ is a free parameter instead.

The parameters $x_{PS}$ and 
$y_{PS}$ are connected with the mixing between $PV$ and
$PS$ channels. 
The representations ${\underline{8}_F}$ (for ${PV}$) 
and ${\underline{8}_D}$ (for ${PS}$) have the same parity and 
charge conjugation 
and may therefore naturally mix, $x_{PS}\neq 0$.
The two $\underline{27}$ representations 
have opposite charge conjugation; the zero hypercharge sectors cannot 
mix if isospin is a 
good symmetry, while the $Y$=$\pm 1$ terms may be mixed with {\sl opposite} 
mixing angles $y_{PS}$ (this is an SU(3) violating effect: SU(3)
symmetry requires {\sl equal} mixing angles for any $Y$ value). We required 
in the fit $|y_{PS}| \leq |x_{PS}|$.

We have to face the problem of enforcing 
orthogonality between the resonant $\underline{8}$ and the non--resonant 
$\underline{27}$ channels. These would be orthogonal in the 
SU(3) symmetric limit, but they are not. For the $PS$ channels, the scalar
multiplet is incomplete and therefore the orthogonality is badly broken.
Even for the $PV$ channels the cancellations that would give orthogonality
do not actually take place, since
we included in the rescattering coefficients the kinematical factors. 
For $D^0$ Cabibbo allowed decays one has 
$\sum_{h'' k''} \tilde{c}_{h'' k''} \tilde{d}_{h'' k''} \simeq 0.16$ and
$\sum_{h' k'} c_{h' k'} d_{h' k'} \simeq 0.008$.
 
The difficulty may be nicely overcome taking advantage of the mixing between 
$PV$ and $PS$ final states. The orthogonality requirement 
\eqn\eORTOG{
\sum_{h' k'} c_{h' k'} d_{h' k'} + x_{PS}\,y_{PS}\,
\sum_{h'' k''} \tilde{c}_{h'' k''} \tilde{d}_{h'' k''} = 0\ }
establishes a relation between $x_{PS}$ and $y_{PS}$,
so that only one of them remains as a free parameter. 
The best fit values are $x_{PS} \simeq 0.25$ and 
$$y_{PS} = \cases { \mp 0.20,& for $D$ Cabibbo allowed (doubly--forbidden)
decays; \cr
\phantom{+}0.00,& for $D_s$ Cabibbo allowed and $D$ first--forbidden decays; \cr
+0.19,& for $D_s$ first--forbidden decays. \cr}$$

We briefly recall now the aspects of \rNOI ~that are not modified in the 
present approach. For the evaluation of the weak decay amplitudes 
${\cal A}_{\rm w}$ we use the factorization approximation and a pole model
for the form factors, as in eqs. \eFACTOR , \eAXIAL . The weak vector 
charges are
assumed SU(3) symmetric: their value, 0.79, is taken from the experimental
results for $D \to K e \nu$. For the axial charges we allow some SU(3) breaking,
and let them vary in the range $0.8 \div 0.9$ independently. 
The decays to final states including $\eta$ or $\eta^{'}$ mesons have
been treated following the approach of D'yakonov and Eides 
\ref\rETAETAP{D.I. D'yakonov and M.I. Eides, Sov.Phys.JETP
54 (1981) 232 \semi see also I. Halperin, Phys. Rev. D 50 (1994) 4602.}: 
the $\eta$--$\eta^{'}$ mixing angle is therefore fixed to -10$^\circ$.
For the decays to
$PP$ and $PV$ channels we also consider the contribution from annihilation
(or $W$--exchange) diagrams:
the relevant matrix elements of the divergences of weak currents
are given in terms of two parameters to be
fitted, $W_{PP}$ and $W_{PV}$, with \rNOI 
\eqn\eDIVER{\eqalign{
       <K^-\pi^+|\partial^\mu(V_s^d)_\mu|0> =&
        \;i\,(m_s-m_d)\,{M_D^2 \over f_D}\,W_{PP}, \cr
       <K^-\rho^+|\partial^\mu(A_s^d)_\mu|0> =&
   -\, (m_s+m_d)\,{2\,M_\rho \over f_D}\,\epsilon^*\cdot p_K\,W_{PV}.\cr}}
The final state interactions for the $PP$ channels are dominated by the 
scalar resonances. Only one of them, the strange $K^*_0(1950)$, 
has been observed 
\ref\rASTON{D. Aston {\it et al}., Nucl. Phys. B 296 (1988) 493.}
in the interesting mass region.
In \rNOI ~we assumed the existence of a nearby isovector resonance $a_0$ 
and we estimated its mass from
the equispacing formula
\eqn\eEQUISP{M_{a_0}^2 = M^2_{K^*_0}-M^2_K+M^2_\pi .}
In the fit we allowed the mass, width and branching ratio in the $K\pi$ channel
\foot{Actually, the parameter to be fitted is the ratio 
$r=g_{818}/g_{888}$, where $g_{818}$ is the 
SU(3) invariant coupling  of the octet of scalar resonances 
to a singlet and an octet of pseudoscalar mesons and $g_{888}$ is the coupling
to two pseudoscalar octets \rNOI . Nonet symmetry corresponds to $r$ = 1. 
The branching ratio is a quadratic function of $r$.}
of the $K^*_0$ resonance to vary
within the experimental bounds. 
From their best fit values (1930
MeV, 300 MeV and 63.5\%, respectively) we get $M_{a_0}$ =
1870 MeV and $\Gamma_{a_0}$ = 299.4 MeV.

In the nonstrange isoscalar case, only relevant for $D^0$
first--forbidden decays, the situation is complicated
by the possibility of singlet--octet mixing of not yet established 
resonances. 
The number of parameters (mixing angles, masses and
coupling constants) is {\sl a priori} quite large. We imposed
the decoupling of the higher mass resonance from the $\pi \pi$ channel, 
which together with the requirement of orthogonality reduces the number of
 new parameters to two: the mixing angle 
\foot{Denoting by $|f_0 \rangle$ the lower mass state, we define
$|f_0\rangle =  \sin \phi \; |f_8\rangle + \cos \phi \;
|f_1\rangle\;, $ $|f_0^{'}\rangle = -\cos \phi \; |f_8\rangle + \sin \phi \;
|f_1\rangle.$}
 $\phi$ and the
difference $\Delta^2=m_{f_0'}^2-m_{f_0}^2$ of the mass squared 
(see \rNOI ~for details). Using the fitted parameters, the masses and widths 
of the two scalar isoscalar resonances are ($M_{f_0},\;\Gamma_{f_0}$) =
(1789, 354) MeV and  ($M_{f_0^{'}},\;\Gamma_{f_0^{'}}$) = (2127, 328) MeV.

We performed a least square fit with 15 parameters to the 49 data points 
or experimental bounds for the branching ratios.
The results are presented in Tables 1 to 4, together with predictions for
the channels not yet measured. 
The values of the eleven  
parameters already used in the previous fits are now: $\xi=0.015$, 
$a_{cu}=a_{cd}=0.9$,
$a_{cs}=0.8$, $W_{PP}=-0.269$, $W_{PV}=0.270$,  
$M_{K^*_0}=1930\;{\rm MeV}$, $\Gamma_{K^*_0}=300\;{\rm MeV}$, $r=-0.86$,
$\phi=47.7^\circ$, $\Delta=1149.4\;{\rm MeV}$ and $\delta_8^{I=0}= 236.5^\circ$.
In \rNOI ~the axial charges were $a_{cu}=a_{cd}=1.0$ and
$a_{cs}=0.59$, while the other parameters are not changed much. 
We list again the four ``new" parameter values:
$\delta_{27}(m_D)=47.4^\circ$, $\delta_{27}(m_{D_s})=59^\circ$, $a_S=0.390$ 
and $x_{PS}=0.249$. The values of decay constants, quark masses and resonance
parameters not explicitely mentioned are identical to the values given in
\rNOI .

The total 
$\chi^2$ is 70.3 (of which 6.2 from two Cabibbo doubly--forbidden decays and
two decays to $PS$ final states, not included in the previous fits). 
In ref. \rNOI , $\chi^2$ was 90 for 45 data points and 11 parameters.
A more detailed comparison of the two fits is shown in Table 5. We note that the
most remarkable improvement occurs for the $D^+ \to PV$ decays: it is
mainly due to rescattering in the exotic $I={3 \over 2}$ channel, that 
is the only rescattering effect present in the Cabibbo--allowed $D^+$
decay amplitudes. 
The worst single point in the fit of ref. \rNOI , the branching ratio
B$(D^+ \to \overline{K}^{*0} \pi^+)$ (that was B$_{th}=  0.64\;\%$ versus
B$_{exp} = 2.2 \pm 0.4\;\%$), is now fitted quite well, B$_{th}= 2.47\;\%$.
This overcompensates the slightly worse fit for the decay $D^+ \to K_S \rho^0$: 
B$_{th}= 5.60\;\%$ now ($5.28\;\%$ in \rNOI ) 
versus B$_{exp} = 3.3 \pm 1.25\;\%$. The greater freedom provided by
the presence of the new parameters $\delta_{27}$ allows the reduction of
the SU(3) breaking in the axial constants $a_{cu}=a_{cd}$ and 
$a_{cs}$, that we imposed not to differ by more than 0.1 in this work.
It also allows an apparently minor change in the annihilation parameters and
in the parameter $\xi$, which now happens to be small and positive: this
has the effect of improving considerably the success of the fit also for
the decay $D^+ \to K_S \pi^+$: B$_{th}= 1.35\;\%$ (it was $1.08\;\%$) versus 
B$_{exp} = 1.37 \pm 0.15\;\%$. A considerable improvement also 
occurs for the Cabibbo forbidden decay 
$D^+ \to  K^+ \overline{K}^{*0}$: B$_{th}= 0.38\;\%$ (it was $0.25\;\%$) 
versus B$_{exp} = 0.51 \pm 0.10\;\%$.

Concerning the decay rates of $D_s^+$ and $D^0$, the quality of the present fit
is comparable to the fit in ref. \rNOI . In particular, for
$D^0 \to \overline{K}^{*0} \eta$ and $D_s^+ \to \rho^+ \eta^{'}$ the results
are still unsatisfactory
 (more than three standard deviations lower than the data points). 
Neither annihilation contributions, nor
 final state interactions were  present for channels with positive
$G$--parity and $I=1$, like $\rho^+ \eta^{'}$, in \rNOI . In this fit the exotic
rescattering affects these channels, giving for instance a nonzero branching 
ratio
for the decay $D_s^+ \to \omega \pi^+$; however, it only slightly lowers
(going in the wrong direction)   
the theoretical prediction for $D_s^+ \to \rho^+ \eta^{'}$.
It might be possible
to attribute the discrepancy
\foot{The large branching
ratio for $D_s^+ \to \rho^+ \eta^{'}$ is difficult to reproduce in many a
model, see also
\ref\rKAEDING{ I. Hinchliffe, T.A. Kaeding, preprint LBL-35892, 
hep-ph/9502275.}.}
 to an annihilation contribution, not taken into account here, through 
the glue components in $\eta^{'}$ and $\eta$
\ref\rBALL{P. Ball, J.M. Fr\`ere and M. Tytgat, preprint CERN-TH-95-220,
hep-ph/9508359.}.

Two out of four data points not included in the fit of \rNOI ~are very well
fitted, but the predictions for the other two are not equally satisfactory.
The amplitude for the decay $D^0 \to f_0 K_S$ is colour suppressed and is
further decreased by the rescattering effects in our model: 
the theoretical value is therefore smaller than the experimental datum.
The doubly--forbidden decay $D^+ \to K^+ \phi$ can only proceed through
annihilation or rescattering: also in this case, the theoretical value is
considerably lower than experiment. It should be noted, however, that
recent data from E791 collaboration 
\ref\rPUROHIT{see M.V. Purohit in Proceedings of the XXVII International
HEP Conference, Glasgow, july 1994, p. 479.} do not observe a signal in this
channel and establish an upper bound slightly less than the central value of 
E691 
\ref\rANJOS{J.C. Anjos et al., Phys.Rev.Lett. 69 (1992) 2892.}, reported in
Table 1. 

As to the predictions for not yet measured decay branching ratios, the
largest among them refers to the Cabibbo first--forbidden decay
$D^+ \to \overline{K}^0 K^{*+}$. The decay amplitude is colour favoured in
this case, and it has a small interfering annihilation contribution instead
of the larger, although colour suppressed,
contribution present in Cabibbo allowed $D^+$ decays. The same is true for
the process $D^+ \to \overline{K}^0 K^+$. The rescattering effects decrease
the decay rate for $\overline{K}^0 K^+$ (which is in very good agreement with
experiment) and increase instead the rate for $\overline{K}^0 K^{*+}$.
The next bigger prediction, for B($D_s^+ \to K^0 \rho^+$), deserves a
similar comment: it is also increased ($\sim 20\%$) by rescattering effects.
Among Cabibbo doubly--forbidden decays, we predict the largest branching 
fractions ($\sim 5\;10^{-4}$) for the decays $D^+ \to K^{+(*)} \pi^0$.
A check for the assumption we made on the scalar particles will be the
observation of decays with $a_0(980)$ production. The largest prediction 
for not yet observed $PS$ decay channels is B($D^+ \to  a_0^+ K_S) = 0.32\;\%$. 

We will not present here the predictions for CP violating decay asymmetries, 
that depend strongly on the rescattering phases: therefore, they remain 
similar to those previously published  
\foot{We remark that all the asymmetries reported in Table V of 
ref. \rNOI ~in correspondence to $D^0$ decays have a wrong sign.
The signs for the charged $D$ decay asymmetries are correct.}
for the $PP$ final states, and differ appreciably in some cases 
for the $PV$ channels. The largest asymmetries ($\sim -3\;10^{-3}$) are now
predicted in the decays $D^+ \to \rho^+ \eta$ and $D^0 \to \omega \eta^{'}$:
they are entirely due to exotic rescattering, and were therefore zero in
\rNOI . The branching ratios of these decays are however small, so that the 
best candidate should be given by the decays $D^+ \to \rho^0\pi^+$ and
$D^- \to \rho^0\pi^-$, the predicted asymmetry being approximately 
$- 2\;10^{-3}$. 
 
A considerable interest has been recently devoted to the interplay
of $D^0 - \bar{D}^0$ mixing and doubly Cabibbo forbidden amplitudes in the
time evolution for $D^0$ decays
\ref\rBROWDER{G. Blaylock, A. Seiden and Y. Nir, Phys.Lett. B 355
(1995) 555 \semi
L. Wolfenstein, Phys.Rev.Lett. 75 (1995) 2460 \semi
T. Browder and S. Pakvasa, preprint UH 511-828-95, hep-ph/9508362.}. 
Particular attention has been given to a term proportional to 
$\Delta M$ and providing linear correction to the exponential 
decay, present as a consequence of $CP$ violation and/or final state
interactions, as a possible source of information on ``new physics". 
A term proportional to $\Delta \Gamma$ is also present.
The short distance contributions predicted by the standard model are very 
small for both $\Delta M$ and $\Delta \Gamma$ 
\ref\rMIXING{see for instance M. Lusignoli, G. Martinelli and A. Morelli,
Phys. Lett. B 231 (1989) 147.}.
It was suggested that the mixing may be dominated by long distance (hadronic)
contributions 
\ref\rLONGDIST{L. Wolfenstein, Phys. Letters 164 B (1985) 170 \semi
J.F. Donoghue, E. Golowich, B.R. Holstein and J. Trampetic,
Phys.Rev. D 33 (1986) 179.}
that could result in mixing parameters $x=\Delta M/\Gamma$ and 
$y=\Delta \Gamma/(2\;\Gamma)$ as large as $10^{-2}$, although this was 
later criticized
\ref\rGEORGI{H. Georgi, Phys. Lett. B 297 (1992) 353.}.

In our model, we can make an estimate of the long distance contribution to
$\Delta \Gamma$ coming from the two--body states that we included in our fit.
This quantity should vanish in the SU(3) limit, through an exact cancellation
of the contribution of Cabibbo allowed and doubly--forbidden transitions
with the contribution of once--forbidden decays \rLONGDIST . 
In the presence of SU(3) breaking the cancellation is however not complete. 
As a consequence, our prediction for $\Delta \Gamma$
is subject to a large uncertainty; on the other hand,  
it is to be noted that the prediction is independent on the rescattering, 
provided that, as we impose, the sum of
the branching ratios remains the same before and after rescattering corrections.

We have ($|\bar{D}^0\rangle =CP\;|D^0\rangle $)
\eqn\eDGAMMA{
    \Gamma_{12} \;=\; \sum_{|f>} {\cal B}^*(D^0 \to f) \; 
{\cal B}\,(\bar{D}^0 \to f)\;\simeq (1.5 + i 0.0014)\,10^{-3} \Gamma_{D^0}}
In \eDGAMMA  ~the sum has been approximated including only 
the contributions of $PP$ (2/3), $PV$ (1/3) and $PS$ ($\sim$0)
final states.  
Note that the contribution  to $\Gamma_{12}/\Gamma_{D^0}$ coming from
Cabibbo first--forbidden decays alone is $35.2\;10^{-3}$, showing that the
SU(3) cancellation is still rather effective. 
Although larger than the short--distance prediction, our estimate is 
much smaller than the present \rPDG ~experimental bound  $|y|=|\Gamma_{12}|\;/\;
\Gamma_{D^0} \leq 0.08$.
The positive sign of the real part of $\Gamma_{12}$ means
(if taken seriously) that the shorter lifetime
state, $D^0_S$, decays dominantly into $CP$--even final states, similarly
to the neutral $K$ mesons.
\vskip 18 pt
M.L. acknowledges an interesting dicussion with L. Maiani, S. Malvezzi and 
D. Menasce.

\listrefs
\vfill\eject

\begintable
$f_i$ | B$_{exp}(D^+ \rightarrow f_{i})$ | B$_{th}$ ||
$f_i$ | B$_{exp}(D^+ \rightarrow f_{i})$ | B$_{th}$ \cr
$K_{S} \pi^{+}$            |$1.37 \pm 0.15$  | $1.35$  ||
$\pi^+ \pi^0 $             | $0.25 \pm 0.07 $ | $0.19$ \cr
$K_{L} \pi^{+}$            | $-$  | $1.70$  ||
$\pi^+ \eta $              | $0.75 \pm 0.25 $ | $0.34$ \cr
$\overline{K}^{*0} \pi^+ $ |$2.2 \pm 0.4 $ | $2.47$ ||
$\pi^+ \eta^{'}$      | $< 0.9 $ | $0.73$ \cr
$K_{S} \rho^+  $           |$3.30 \pm 1.25 $  | $5.60$ ||
$\overline{K}^{0} K^+ $    | $0.78 \pm 0.17$ | $0.81$ \cr
$K_{L} \rho^+  $           | $-$  | $6.30$ ||
$\rho^0 \pi^+ $            | $< 0.14$  | $0.13$ \cr
$a_0^+\;K_{S}  $  |  $-$  |  $0.32$ ||
$\rho^+ \pi^0 $            | $ - $     | $0.44$ \cr
$a_0^+\;K_{L}  $  |  $-$  |  $0.24$ ||
$\rho^+ \eta $             | $< 1.2$   | $0.013$ \cr
$K^+ \pi^0 $  | $-$  | $0.056$ ||
$\rho^+ \eta^{'} $    | $< 1.5$ | $0.12$ \cr
$K^+ \eta $  | $-$  | $0.018$ ||
$\omega \pi^+ $            | $< 0.7$ | $0.019$ \cr
$K^+ \eta^{'} $  | $-$  | $0.031$ ||
$\phi \pi^+ $              | $0.67 \pm 0.08$ | $0.61$ \cr
$K^{*0} \pi^+ $  |  $-$  | $0.019$ ||
$\overline{K}^0 K^{*+} $   | $ - $ | $ 1.71 $ \cr
$K^{*+} \pi^0 $ | $-$ | $0.048$ ||
$\overline{K}^{*0} K^+ $   | $0.51 \pm 0.10 $ | $0.38$ \cr 
$K^{*+} \eta $ | $-$ | $0.030$ ||
$f_0\;\pi^+  $ |  $-$   |  $0.028$ \cr
$K^{*+} \eta^{'} $ | $-$ | $0.0002$ ||
$a_0^0\;\pi^+ $   |  $-$  |  $0.059$  \cr
$K^+ \rho^0 $  | $-$  | $0.030$ ||
$a_0^+\;\pi^0 $ |  $-$  |  $0.012$  \cr
$K^+ \omega $  | $-$  | $0.021$ ||
$a_0^+\;\eta $ |  $-$  |  $0.074$ \cr
$ K^+ \phi$   | $0.039 \pm 0.022$ | $0.0051$ ||
$K^+ f_0 $ | $-$ | $0.0023$ \cr
 | | ||
$K^+ a_0^0 $ | $-$ | $0.0062$
\endtable
\vskip 0.5 cm
\centerline{TABLE 1}
\vskip 0.3 cm
\noindent
\centerline{Branching ratios for $D^+$ nonleptonic decays.}
\centerline{[Experimental data and 90\% c.l. upper bounds from ref. \rPDG ]}
\vfill\eject
\begintable
$f_i$
| B$_{exp}(D_s^{+} \rightarrow f_{i})$ | B$_{th}$ ||
$f_i$
| B$_{exp}(D_s^{+} \rightarrow f_{i})$ | B$_{th}$  \cr
$K_{S} K^{+}$          |$1.75 \pm 0.35$  | $2.37$  ||
$K^+ \pi^0$            |$-$ | $0.14$  \cr
$K_{L} K^{+}$          | $-$  | $2.09$  ||
$K^+ \eta $            | $-$ | $0.28$ \cr
$\pi^+ \eta $          | $1.90 \pm 0.40 $ | $1.23$ ||
$K^+ \eta^{'} $   | $-$ | $0.44$ \cr
$\pi^+ \eta^{'} $ | $4.7 \pm 1.4$ | $5.39$ ||
$K^0 \pi^+$            |$< 0.7$  | $0.40$  \cr
$\rho^+ \eta $         | $10.0 \pm 2.2$ | $7.49$ ||
$K^{*+} \pi^0$         |$-$ | $0.044$ \cr
$\rho^+ \eta^{'} $| $12.0 \pm 3.0$ | $2.41$ ||
$K^+ \rho^0 $          | $-$ | $0.29$ \cr
$\overline{K}^{*0} K^+$|$3.3 \pm 0.5 $  | $3.96$ ||
$K^{*+} \eta $         |  $-$ | $0.18$ \cr
$K_{S} K^{*+}  $       |$2.1 \pm 0.5 $ | $1.87$  ||
$K^{*+} \eta^{'} $| $-$  | $0.025$ \cr
$K_{L} K^{*+}  $       | $-$ | $2.13$  ||
$K^+ \omega$           | $-$  | $0.15$ \cr
$\phi \pi^+ $          | $3.5 \pm 0.4$  | $4.08$ ||
$K^+ \phi$             | $< 0.25$ | $0.018$ \cr
$\omega \pi^+ $        | $< 1.7$  | $0.26$ ||
$K^{*0} \pi^+ $        |$-$ | $0.29$ \cr
$\rho^0 \pi^+ $        |$< 0.28$  | $0.24$ ||
$K^0 \rho^+$           |$-$ | $1.39$  \cr
$\rho^+ \pi^0 $        |$-$ | $0.24$ ||
$f_0\;K^+  $           |$-$ | $0.069$ \cr
$f_0\;\pi^+  $         |  $1.0 \pm 0.4$ | $1.06$ ||
$a_0^+\;K^0  $        |  $-$  |  $0.003$  \cr
$a_0^+\;\eta $         |  $-$  |  $0.007$ ||
$a_0^0\;K^+  $        |  $-$  | $0.007$ \cr  
$a_0^+\;\eta^{'}  $  |  $-$  |  $0.002$ ||
$ K^{*0} K^+ $  |  $-$  | $0.008$
\endtable
\vskip 0.5 cm
\centerline{TABLE 2}
\vskip 0.3 cm
\noindent
\centerline{Branching ratios for $D_s^+$ nonleptonic decays.}
\centerline{[Experimental data and 90\% c.l. upper bounds from ref. \rPDG ]}
\vfill\eject
\begintable
$f_i$ | B$_{exp}(D^0 \rightarrow f_{i})$ | B$_{th}$ ||
$f_i$ | B$_{exp}(D^0 \rightarrow f_{i})$ | B$_{th}$ \cr
$K^{-} \pi^+ $  |$4.01 \pm 0.14$      |  $4.04$ ||
$\pi^0 \eta$               |$-$ | $0.052$ \cr
$K_{S} \pi^{0}$ |$1.02 \pm 0.13$      | $0.72$ ||
$\pi^0 \eta^{'}$      | $-$ | $0.16$ \cr
$K_{L} \pi^{0}$     | $-$ | $0.53$ ||
$\eta \eta$                |$-$ | $0.088$ \cr
$K_{S}\; \eta$  |$0.34 \pm 0.06$      | $0.42$ ||
$\eta \eta^{'}$       |$-$ | $0.18$ \cr
$K_{L}\; \eta$  | $-$ | $0.31$ ||
$\pi^0 \pi^0$              |$0.088 \pm 0.023$ | $0.110$ \cr
$K_{S}\; \eta^{'}$ | $0.83 \pm 0.15$   | $0.78$ ||
$\pi^+ \pi^-$              |$0.159 \pm 0.012$ | $0.159$ \cr
$K_{L}\; \eta^{'}$ | $-$ | $0.61$ ||
$K^+ K^-$                  |$0.454 \pm 0.029$ | $0.446$ \cr
$\overline{K}^{*0} \pi^0 $ |$3.0 \pm 0.4 $  | $3.49$ ||
$K^0 \KB$                  |$0.11 \pm 0.04$   | $0.098$ \cr
$K_{S}\; \rho^0  $ | $0.55 \pm 0.09 $       | $0.47$  ||
$\omega \pi^0$             | $-$   | $0.014$ \cr
$K_{L}\; \rho^0  $ | $-$ | $0.33$  ||
$\rho^0 \eta$              | $-$   | $0.020$ \cr
$K^{*-} \pi^+ $    |$4.9 \pm 0.6 $   | $4.85$  ||
$\rho^0 \eta^{'}$     | $-$   | $0.008$ \cr
$K^{-} \rho^+  $   |$10.4 \pm 1.3 $  | $11.02$  ||
$\omega \eta$              | $-$   | $0.20$ \cr
$\overline{K}^{*0} \eta $ |$ 1.9 \pm 0.5 $ | $0.37$ ||
$\omega \eta^{'}$     | $-$   | $0.0001$ \cr
$\overline{K}^{*0} \eta^{'} $ | $ < 0.11 $ | $ 0.004 $ ||
$\phi \pi^0$               | $-$   | $0.11$ \cr
$K_{S}\; \omega  $ |$1.0 \pm 0.2 $  | $0.88$ ||
$\phi \eta$                | $-$   | $0.090$ \cr
$K_{L}\; \omega  $ | $-$ | $0.80$ ||
$K^{*0} \KB$               | $< 0.08$ | $0.064$ \cr
$K_{S}\; \phi  $   | $0.415 \pm 0.060 $  | $0.40$ ||
$\overline{K}^{*0} K^0 $   | $< 0.15$ | $0.062$  \cr
$K_{L}\; \phi  $   | $-$  | $0.42$ ||
$K^{*+} K^-$               | $0.34 \pm 0.08$ | $0.43$ \cr
$f_0\;K_{S}  $    |  $0.23 \pm 0.10$ | $0.037$  ||
$K^{*-} K^+$               | $0.18 \pm 0.10$ | $0.30$ \cr
$f_0\;K_{L}  $    | $-$ | $0.031$  ||
$\rho^+ \pi^-$             | $-$ | $0.69$ \cr
$a_0^0\;K_{S}  $  |  $-$  |  $0.109$  ||
$\rho^- \pi^+$             | $-$ | $0.57$ \cr
$a_0^0\;K_{L}  $  |  $-$  |  $0.083$  ||
$\rho^0 \pi^0$             | $-$ | $0.12$ \cr
$a_0^+\;K^{-}  $  |  $-$  |  $0.078$ ||
 | | 
\endtable
\vskip 0.5 cm
\centerline{TABLE 3}
\vskip 0.3 cm
\noindent
\centerline{Branching ratios for $D^0$ Cabibbo allowed and first--forbidden 
decays.}
\centerline{[Experimental data and 90\% c.l. upper bounds from ref. \rPDG ]}
\vfill\eject
\begintable
$f_i$ | B$_{exp}(D^0 \rightarrow f_{i})$ | B$_{th}$ ||
$f_i$ | B$_{exp}(D^0 \rightarrow f_{i})$ | B$_{th}$ \cr
$f_0\;\pi^0  $   |  $-$  |  0.0006  ||
$ K^+ \pi^-  $   | $0.031 \pm 0.014$ | $0.033$ \cr
$f_0\;\eta $    | $-$  | $0.004$  ||
$K^{*0} \pi^0 $  |  $-$  | $0.0039$ \cr
$a_0^0\;\pi^0  $   |  $-$   | $0.011$   ||
$K^{*+} \pi^- $ | $-$ | $0.035$ \cr
$a_0^0\;\eta  $   |  $-$  |  $0.015$  ||
$ K^+ \rho^- $  | $-$ | $0.025$ \cr
$a_0^+\;\pi^{-} $ | $-$  |  $0.003$  ||
$K^{*0} \eta $ |  $-$ | $0.009$  \cr
$a_0^-\;\pi^{+} $ | $-$  |  $0.070$ ||
$K^{*0} \eta^{'} $ | $-$ | $\sim 10^{-5}$ \cr
  | | || 
$a_0^-\;K^+  $  |  $-$  |  $0.004$ 
\endtable
\vskip 0.5 cm
\centerline{TABLE 4}
\vskip 0.3 cm
\noindent
\centerline{Branching ratios for $D^0$ Cabibbo first-- and doubly--forbidden 
decays.}
\centerline{[Experimental data from ref. \rPDG ]}
\vskip 2.5 cm
\begintable
  Decays   | \# data  | $\chi^2$ (ref. \rNOI ) | $\chi^2$ (This work)  \cr
$D^+ \to PP$ | 5  | 9.56  |  5.34  \cr
$D^+ \to PV$ | 8  | 29.55  |  8.46  \cr
$D_s^+ \to PP$ | 4  | 8.79 |  7.10  \cr
$D_s^+ \to PV$ | 8  | 15.35 |  17.64 \cr 
$D^0 \to PP$ | 8  | 8.44  |  8.43  \cr
$D^0 \to PV$ |12  | 18.35  |  17.17  
\endtable
\vskip 0.5 cm
\centerline{TABLE 5}
\vskip 0.3 cm
\noindent
\centerline{Comparison of our results with the fit of ref. \rNOI .} 
\centerline{Only Cabibbo--allowed and first--forbidden decays are included.}
\vfill\eject

\bye